\documentclass{optica-article}

\journal{opticajournal} 

\articletype{Research Article}

\usepackage{lineno}

\begin{document}

\title{On fine alignment of transmitted beams for TianQin with far-field wavefront error}

\author{Dezhi Wang,\authormark{1} Xuefeng Zhang\authormark{1,*} and Hui-Zong Duan,\authormark{1,*}}

\address{\authormark{1}MOE Key Laboratory of TianQin Mission, TianQin Research Center for Gravitational Physics $\&$ School of Physics and Astronomy, Frontiers Science Center for TianQin, Gravitational Wave Research Center of CNSA, Sun Yat-sen University (Zhuhai Campus), Zhuhai 519082, China}

\email{\authormark{*}zhangxf38@sysu.edu.cn} 
\email{\authormark{*}duanhz3@sysu.edu.cn}

 
\begin{abstract*} 
TianQin is a proposed space-based gravitational wave detector mission that employs inter-satellite laser interferometry. Suppressing measurement noise and achieving high sensitivity require accurate alignment of multiple onboard interferometers after laser link acquisition. However, due to huge armlengths and varying point-ahead angles, the fine alignment of the transmitted beams can be particularly challenging, which needs to take into account both received laser power and far-field wavefront errors. To tackle this issue for TianQin which has small point-ahead angle variations, we propose an efficient alignment strategy that relies on finding the maximum-intensity direction of the transmitted beam as the alignment reference. The direction can be estimated through a quatrefoil scan of the local transmitted beam and the corresponding intensity measurement from the remote satellite. Under TianQin's fixed-value compensation of the point-ahead angles, simulation results reveal that the proposed strategy is capable of aligning the transmitted beams within 20 nrad from the mean value of the point-ahead angles, while the tilt-to-length coupling associated with far-field wavefront error can meet the requirement given a transmitted beam aberration of $\lambda/40$ RMS. 
\end{abstract*}

\section{Introduction}
\begin{figure}[htb]
	\centering
	\includegraphics[width=0.7\textwidth]{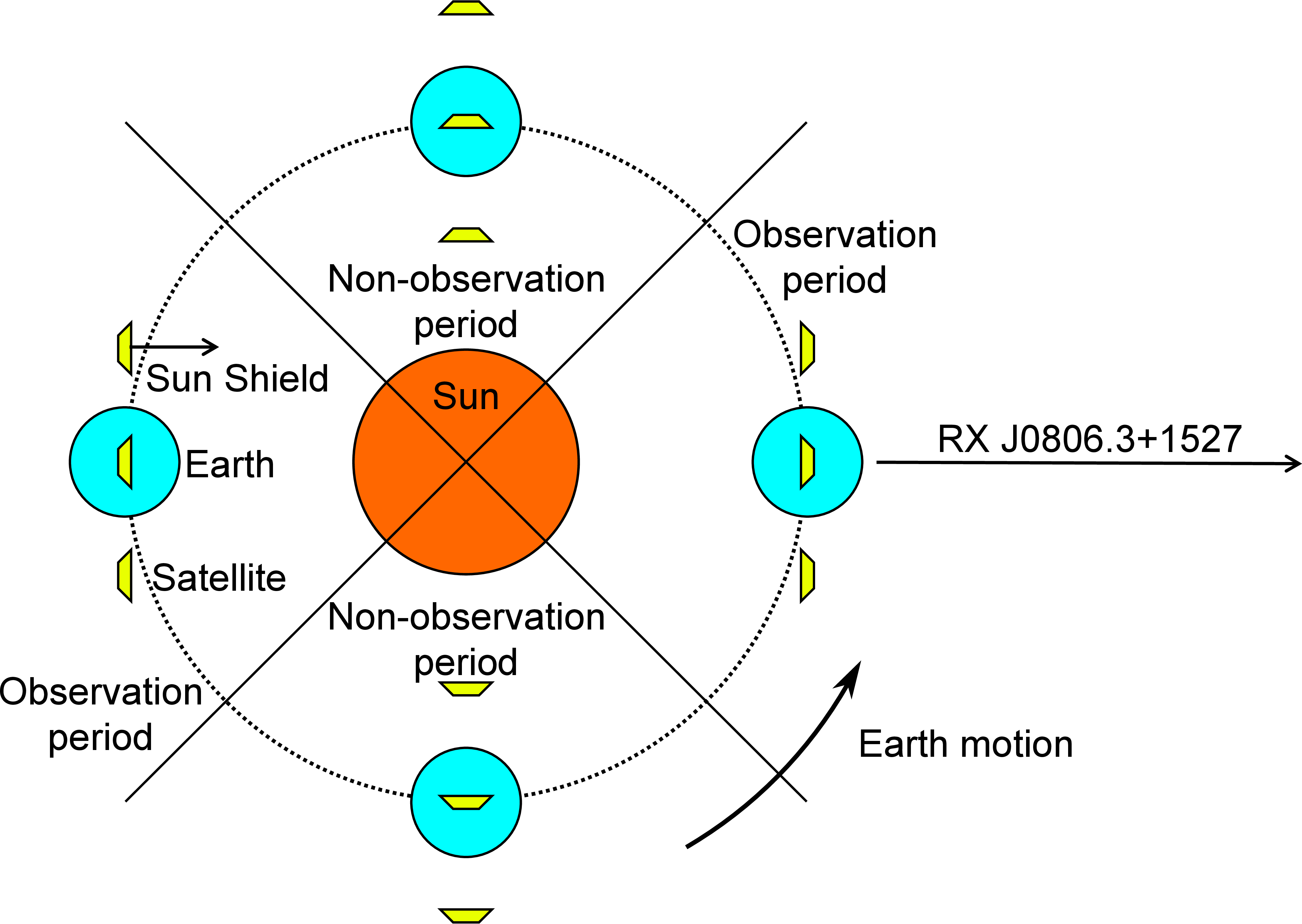}
	\caption{A schematic diagram of observation periods and non-observation periods for TianQin. The constellation is composed of three satellites surrounding the earth. The orientation of the constellation plane remains nearly fixed and points to the reference source RX J0806.3+1527. During an observation period, the sun shield of satellite keeps parallel to the constellation plane and between two observation periods, the satellites flip themselves.}
	\label{fig:3+3}
\end{figure}

TianQin is a space-based gravitational wave detection mission utilizing a three-satellite constellation deployed in geocentric orbits and arranged as an approximately equilateral triangle \cite{Luo2016}. The orbit radius is $10^8$ m, and the inter-satellite arm length is $\sqrt{3} \times 10^8$ m. The mission adopts RX J0806.3+1527 as its reference gravitational wave source, requiring the constellation plane to maintain an orientation nearly perpendicular to the ecliptic plane while constantly pointing toward this source. As Earth orbits the Sun, the solar aspect angle relative to TianQin's constellation plane varies periodically. When this angle becomes too small, the interferometric measurements are compromised, necessitating an operational cycle that alternates between 3-month observation periods and 3-month non-observation periods ("3+3" mode, see Fig. \ref{fig:3+3})\cite{xuefengzhang2018}.

The main scientific payload of TianQin is the Movable Optical Sub-Assembly (MOSA), an integrated system comprising an inertial sensor, an optical bench (OB), and a telescope \cite{Weise2006}. The inertial sensor houses a precisely shielded test mass (TM) maintained in free-fall along the interferometer axis, serving as the reference for gravitational wave measurements. By measuring variations in the separation distance between the centers of mass (CoM) of TMs on board each satellite pair via laser interferometry, TianQin is able to detect gravitational waves in the 0.1 mHz to 1 Hz frequency band. The optical bench integrates multiple interferometers to achieve picometer-level ranging accuracy, while simultaneously providing satellite attitude data through Differential Wavefront Sensing (DWS) \cite{morrison1994automatic}. The afocal telescope adjusts the diameters of the incoming and outgoing beams as well as their direction angles.

Each TianQin satellite carries two MOSAs with a nominal 60$^\circ$ angle between them to maintain laser links with both neighboring spacecrafts simultaneously. Due to the orbital dynamics, the angle exhibits fluctuation, i.e., breathing angle. The MOSA features an active one-degree-of-freedom pointing adjustment capability relative to the spacecraft platform within the constellation plane to compensate for the breathing angle. The pointing control of MOSA is implemented by the drag-free pointing control (DFPC) system, which is designed to maintain both TMs in free fall along the interferometric axis, actively control their other degrees of freedom, and keep the satellite platform precisely tracking their motion \cite{fang2024payload}. The DFPC system also regulates the off-plane orientation of MOSA by maneuvering the spacecraft. The MOSA orientation specification requires a pointing bias of less than 10 nrad and jitter of less than 10 nrad/$\sqrt{\mathrm{Hz}}$.

To simplify the process of assembling and alignment, TianQin employs a split-interferometry architecture that decomposes the TM-to-TM displacement into three distinct optical path segments: the displacement between the local TM and local OB, measured by the local test mass interferometer (TMI); the inter-satellite displacement between OBs, measured by the inter-satellite interferometer (ISI); and the displacement between the remote OB and remote TM, measured by the remote TMI \cite{otto2015}. This segmented approach isolates different noise sources while the end-to-end measurement can be reconstructed.

The angle between a MOSA's received and transmitted beams is defined as point-ahead angle (PAA), which is induced by the relative lateral motion between satellites and the finite speed of light. The PAA varies over time due to orbital dynamics. In Ref. \cite{Wang_2024}, we calculated TianQin's PAA characteristics and proposed a customized compensation strategy. The in-plane component of PAA exhibits a constant bias of 23.07 $\mu$rad with a variation range of 55 nrad, while the off-plane component shows negligible bias with variations of $\pm$10 nrad. Consequently, we proposed a fixed PAA compensation strategy that only the in-plane PAA bias should be compensated in a stationary manner during science mode. The pointing-ahead angle mechanism (PAAM) utilizes a steering mirror mounted in the outgoing beam path to adjust the beam pointing direction and compensate for PAA.

 The displacement induced by gravitational waves is expected at the picometer level, which requires the total noise to remain below 1 pm/$\sqrt{\mathrm{Hz}}$ across the measurement bandwidth. After suppressing laser frequency noise and clock noise through time-delay interferometry (TDI) \cite{tinto2021time, hartwig2021clock, bayle2023unified, zheng2023doppler}, tilt-to-length (TTL) coupling noise becomes the dominant noise source \cite{paczkowski2022postprocessing}. TTL noise is caused by angular or lateral displacements of the beam relative to the optical system and is present in both the ISI and TMI. The primary factors affecting TTL noise are the optical path geometry and the beam's physical properties, leading to its classification into geometric and non-geometric TTL noise components \cite{hartig2022geometric, hartig2023non}. Under the DFPC system, the MOSA's angular jitter remains sufficiently small that the TTL noise can generally be approximated as a linear product of the TTL coupling coefficients and the corresponding jitter amplitudes.

 In general, space-based gravitational wave detectors suppress the TTL noise via three approaches: design, realignment, and subtraction \cite{paczkowski2022postprocessing, wanner2024depth}. The first approach reduces TTL coupling via optimize the design of optical path, such as incorporating imaging systems in the optical path \cite{trobs2018reducing, chwalla2020optical}. The second method tunes and realigns optical elements in-orbital to diminish TTL coupling. The mission LISA Pathfinder has demonstrated that additional noise suppression can be achieved through precision alignment of the optical configuration \cite{wanner2017preliminary}. The subtraction method involves calibrating TTL coupling coefficients via a linear noise model using science measurement data and jitter measured by DWS, allowing for the estimation and subtraction of TTL noise from scientific measurements \cite{paczkowski2022postprocessing, houba2022lisa,houba2022optimal, george2023calculating, wang2025postprocessing}. Ref. \cite{wang2025postprocessing} has proposed that, to meet TianQin's requirements, the TTL coupling coefficient must remain below 3 pm/nrad with a rate of change (ROC) under 0.1 pm/nrad/day, ensuring residual noise stays within acceptable limits.

The paper primarily focuses on alignment strategies for the ISI, with particular emphasis on transmitted beam alignment. In the ISI, astigmatism and defocus $d$ in the transmitted beam with aperture diameter $D$ can induce non-geometric TTL noise, which can be calculated by
\begin{equation}
    \delta x = \frac{1}{32}\left(\frac{2 \pi}{\lambda}\right)^2 d D^2 \beta_0 \delta \beta
\end{equation}
as proposed in Ref. \cite{winkler1997truncated}, showing that the angular deviation $\beta_0$ between the remote satellite and the beam axis can induce the TTL noise combined with jitter $\delta \beta$. Therefore, proper alignment can suppress such noise in the ISI. Ref. \cite{bender2005wavefront, sasso2018coupling, ming2021analysis, xiao2023analysis} demonstrated through analysis of transmitted beams with finite-order Zernike aberrations that a stationary point exists in the far field where such TTL noise becomes negligible. They also derived analytical formulas for the offset between the stationary point and the beam axis. 

However, in practice, the real beam axis and the actual aberrations can hardly be determined in orbit, making the localization of the stationary point challenging. A straightforward alternative is combining TTL calibration with tuning PAAM to suppress TTL coupling effect. However, accurate calibration of TTL coefficients requires long measurement duration (about 1 day) and ground-based postprocessing \cite{paczkowski2022postprocessing, george2023calculating, wang2025postprocessing}, resulting in prohibitively time consumption. Moreover, TianQin's fixed PAA compensation strategy introduces relative motion between the remote satellite and far-field wavefront, varying TTL coupling coefficients and further complicating the alignment for the transmitted beam \cite{Wang_2024}. 

Given TianQin's 3+3 operation mode, laser links should be reacquired and realigned at the beginning of every observation period. To minimize time consumption and maximize observational efficiency, optimized alignment strategies and procedures must be developed. The primary objective of alignment is to mitigate system noise contributions. There are two main alignment-dependent noise sources in ISI: shot noise related to the received optical power and TTL coupling noise concerned with the distortion of wavefront. In general, minimizing each of them requires distinct alignment state separately, necessitating a trade-off between received power and wavefront quality. Given that TTL coupling noise can be further suppressed through post-processing, we prioritize received power maximization as the primary alignment criterion while staying within post-processing capability for TTL noise subtraction, which consequently imposes constraints on transmitted wavefront aberrations.

The paper is organized as follows: Sec. \ref{sec:simMetheod} introduce TianQin's orbits, establishes some coordinate systems, and details the simulation methodology for inter-satellite beam propagation. Sec. \ref{sec:alignMethod} proposes a calibration methodology for a reference to align the transmitted beam, along with simulations for preliminary validation of effectiveness. Sec. \ref{sec:simulations} presents alignment strategies for TianQin's inter-satellite beam, examines their feasibility, and outlines relative requirements. Sec. \ref{sec:conclusion} provides concluding remarks and suggests directions for future research.

\section{Model setup}
\label{sec:simMetheod}

In this section, we initially introduce TianQin's orbits and establish coordinate systems relevant to our study. Next, we introduce the methods employed throughout this work to simulate the far field and to extract information from it.

\subsection{Orbits, coordinate systems, and PAA compensation}

The orbit of TianQin determines each satellite's motion and constellation's evolution. It is the base for subsequent simulations. We adopt initial orbital elements provided in Ref. \cite{zhang2021effect}, and Table \ref{tab:orb-elements} summarizes them. The coupling effects resulting from attitude maneuvers and the self-gravity perturbation are explicitly included during the simulation of orbit evolution \cite{fang2025arxiv}. Additionally, the geometric offsets between CoMs of satellite and TMs are also incorporated, departing from the conventional assumption of their coincidence.

\begin{table}[htbp]
\caption{Optimized initial orbital elements of TianQin satellites (SC1, 2, 3) in the J2000-based Earth-centered equatorial coordinate system at the epoch 6 June 2004, 00:00:00 UTC for evaluation purposes.}
  \label{tab:orb-elements}
  \centering
\begin{tabular}{cccc}
\hline
 & SC1 & SC2 & SC3 \\
\hline
semimajor axis $a \ \rm{(km)}$ & 100000.0 & 100009.5 & 99995.0 \\
eccentricity $e$ & 0 & 0 & 0 \\
inclination $i \ (^\circ)$ & 74.5 & 74.5 & 74.5 \\
longitude of ascending node $\Omega \ (^\circ)$ & 211.6 & 211.6 & 211.6 \\
argument of periapsis $\omega \ (^\circ)$ & 0 & 0 & 0 \\
true anomaly $\nu \ (^\circ)$ & 30 & 150 & 270 \\
\hline
\end{tabular}
\end{table}

\begin{figure}[htbp]
\centering\includegraphics[width=7cm]{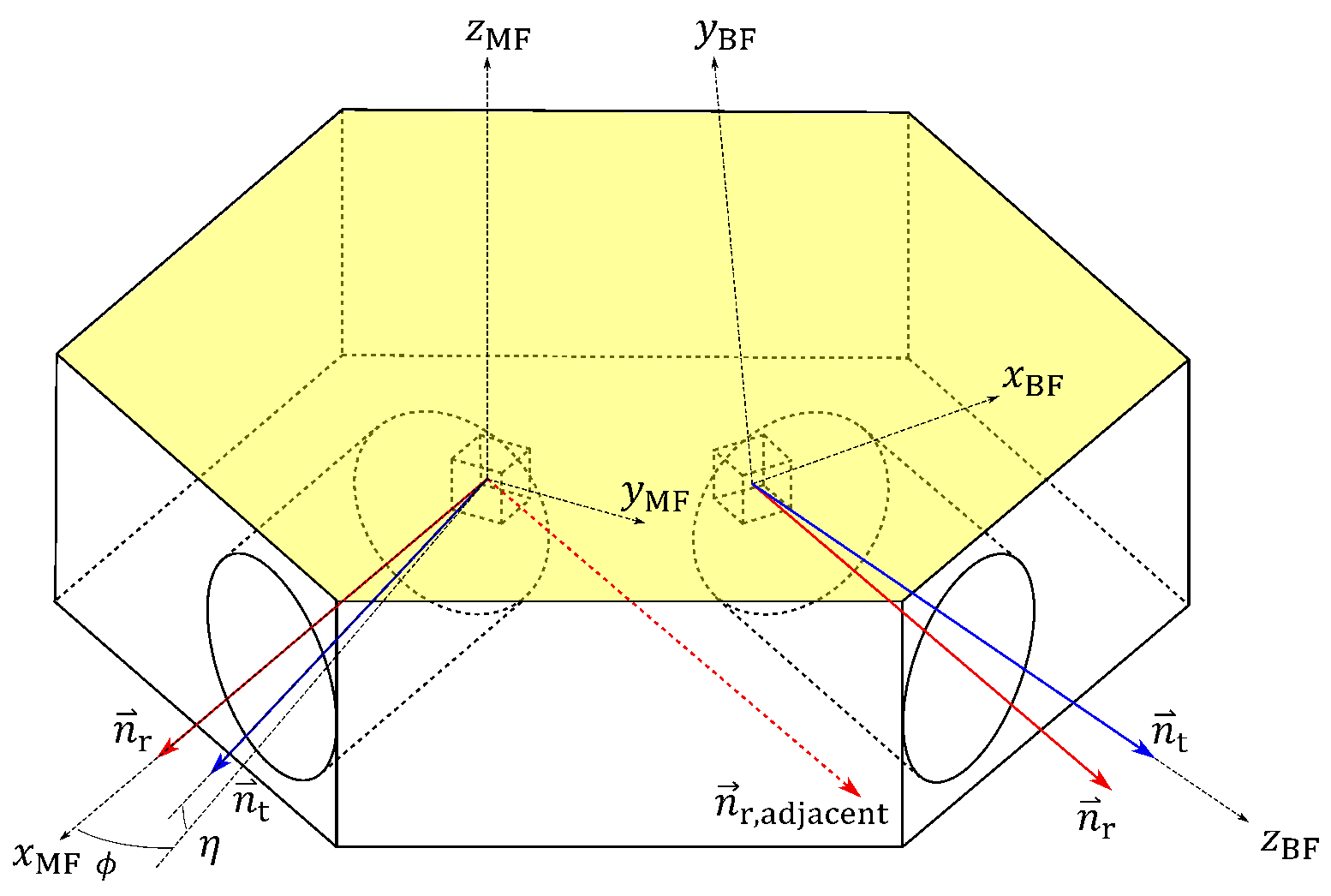}
\caption{A schematic diagram of the MOSA frame (illustrated on the left MOSA) and the beam frame (illustrated on the right MOSA). The vector $\vec{n}_{\rm{r}}$ represents the nominal orientation of each MOSA, which should align normal to the wavefront of the incident light, while the vector $\vec{n}_{\rm{t}}$ depict the nominal direction of each transmitted beam. The angles $\phi$ and $\eta$ denote the in-plane PAA and off-plane PAA, respectively.}
\label{fig:PAAco}
\end{figure}

To characterize PAA and transmitted beam, we have introduced two coordinate systems in Ref. \cite{Wang_2024}: the MOSA frame and the beam frame. The definitions of them are shown in Table \ref{tab:coordinate}, where the unit vector $\vec{n}_{\rm{r}}$ represents the nominal MOSA orientation aligned with the incident beam and $\vec{n}_{\rm{t}}$ represents the direction of the transmitted beam. The configurations of these coordinate systems are illustrated in Fig. \ref{fig:PAAco}. The plane determined by two receiving vectors $\vec{n}_{\mathrm{r}}$ of a single satellite and the TM's CoM is referred to as the local constellation plane (which is also the $x-y$ plane of MOSA frame), the PAA can then be decomposed into in-plane component and off-plane component. To ensure unambiguous identification of individual satellites and MOSAs, three satellites are sequentially labeled 1, 2, and 3. For MOSAs, they inherit index notation from their host satellites and two of them attached to the same satellite are distinguished by the prime notation ($^\prime$). This notation scheme is consistent with one employed in TDI \cite{Zheng_2023}.

\begin{table}[htbp]
\caption{The definitions of the coordinate systems.}
  \label{tab:coordinate}
  \centering
\begin{tabular}{ccc}
\hline
name & MOSA frame & beam frame \\
\hline
origin & nominal position of TM's CoM & nominal position of TM's CoM \\
$x$ & aligning with $\vec{n}_{\rm{r}}$ & aligning with $(\vec{n}_{\rm{r}} \times \vec{n}_{\rm{t}}) \times \vec{n}_{\rm{t}}$ \\
$y$ & aligning with $(\vec{n}_{\rm{r}} \times \vec{n}_{\rm{r, adjacent}}) \times \vec{n}_{\rm{r}}$ & aligning with $\vec{n}_{\rm{r}} \times \vec{n}_{\rm{t}}$ \\
$z$ & aligning with $\vec{n}_{\rm{r}} \times \vec{n}_{\rm{r}}^{\prime}$ & aligning with $\vec{n}_{\rm{t}}$ \\
\hline
\end{tabular}
\end{table}

\begin{figure}
	\centering\includegraphics[width=\textwidth  ]{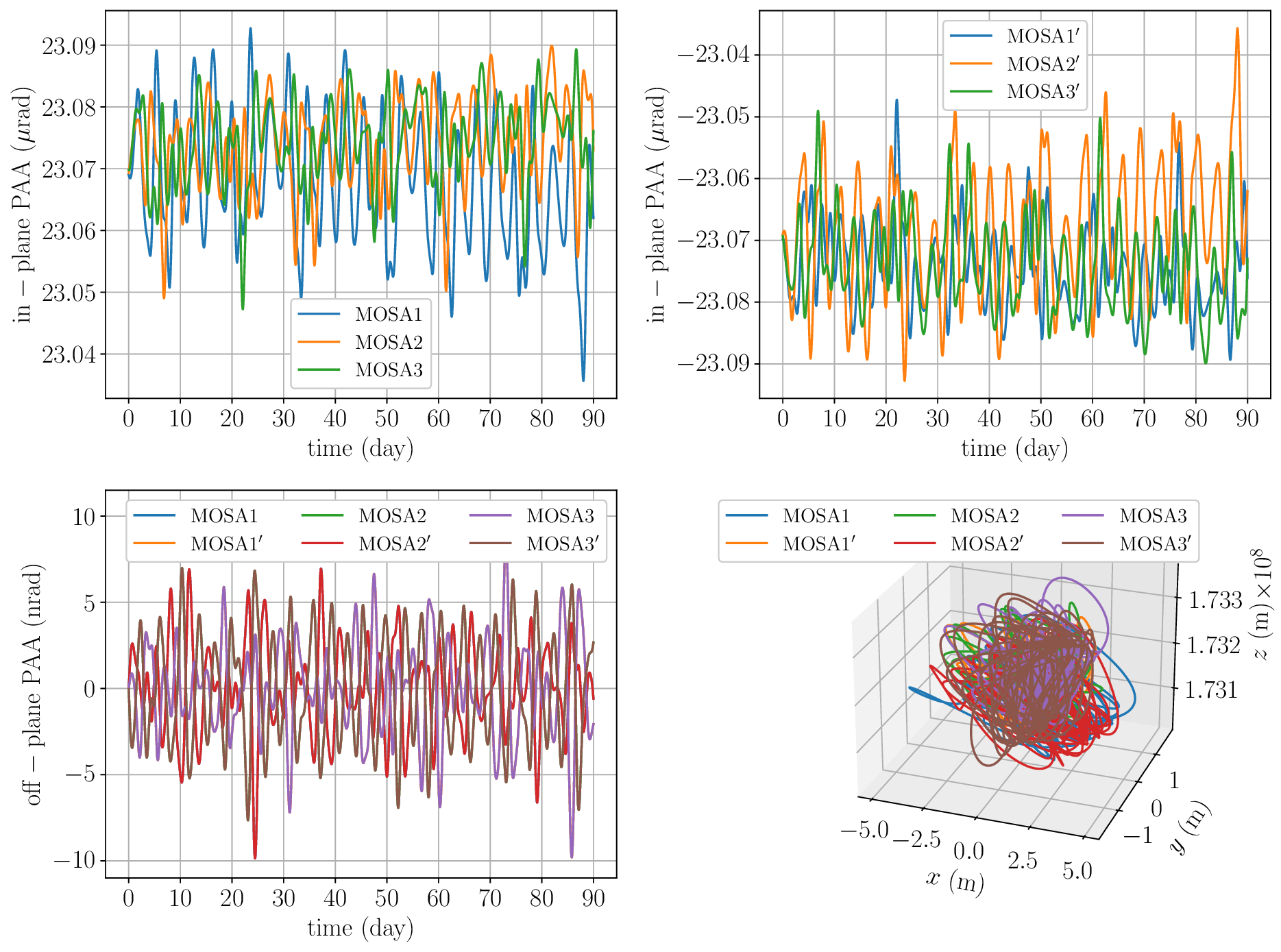}
	\caption{The two upper plots display TianQin's in-plane PAAs, revealing a fixed bias of $23.07 \ \mu \mathrm{rad} $ with a variation of approximately $\pm 25$ nrad. Different sign indicates the different direction of PAA which is induced by constellation rotation. The lower left plot illustrates TianQin's off-plane PAAs in one observation period, showing a near-zero bias with a variation within $\pm 10$ nrad. The lower right plot displays trace of relative motion between each remote satellite and the corresponding beam of each MOSA in the beam frame.}
	\label{fig:paaAndTraces}
\end{figure}

The resulting PAA to each MOSA is shown in Fig. \ref{fig:paaAndTraces}, which demonstrates that the offset between satellite's and TM's CoMs, the coupling effects induced by attitude maneuvers, and the well-constrained self-gravity have insignificant effects on PAA. Therefore, the fixed PAA compensation strategy for TianQin remains feasible. The residual PAA dynamic component after compensation reflects the relative motion between the remote satellite and the transmitted beam, and the strategy makes the transmitted beam points towards the mean position of the remote satellite. The trace of such motion can be derived using the Gaussian beam propagation model (see Ref. \cite{Wang_2024}), and the result for every MOSA is shown in the lower-right plot of Fig. \ref{fig:paaAndTraces}. Each of them can be served as a series of time-sequential target points for the following far-field simulation, enabling rigorous calculation of the optical signal received by the remote satellite.

\subsection{Methodology and parameters for far-field simulation}
\label{subsec:ffsim}

To obtain the far-field information between satellites, we employ the Rayleigh-Sommerfeld diffraction integral
\begin{equation}
	U(P) = \frac{1}{\mathrm{i} \lambda} \int_{\Sigma} U(P_0) \frac{\exp{(\mathrm{i}kl)}}{l} \kappa \mathrm{d}^2 \sigma
	\label{eq:diffractionEquation}
\end{equation}
to calculate the far field, where $\Sigma$ denotes the transmitted aperture region and $P_0$ represents the point within it, while the point $P$ is the far-field target position and $l$ is the distance between them. The scalar field magnitude $U$ is complex, with the intensity given by $|U|^2$ and the phase derived as $\arg(U)$. The wave number $k$ is defined as $ 2 \pi / \lambda$, and for TianQin, the wavelength $\lambda$ of the laser is $1064$ nm. The inclination factor $\kappa$ equals the cosine of the angle between the normal vector of surface $\Sigma$ at point $P_0$ and the vector connecting $P_0$ and $P$.

We simulate the propagation of the transmitted beam in the beam frame. To simplify simulations, we assume the transmitted aperture, with radius $r_0$, is located at the nominal position of TM's CoM, and the $z$ axis is perpendicular to the aperture plane. The transmitted transverse field distribution $U(r, \theta, 0)$ is expressed as $U_0(r, \theta, 0)\cdot\exp[\mathrm{i} k w_{\mathrm{e}}(r, \theta, 0)]$, where the radial amplitude distribution of amplitude is given by a Gaussian profile
\begin{equation}
	U_0(r, \theta, 0) = \sqrt{\frac{2P}{\pi w^2}} \exp \left(-\frac{r^2}{w^2}\right)
\end{equation}
with total power of the transmitted beam $P = 2.72$ W before aperture clipping \cite{Freise2010}. The beam radius $w$ is defined as the position with $1/\mathrm{e}^2$ maximum intensity. To maximize the on-axis intensity in the far field, the beam waist $w$ is optimized to 0.8921$r_0$ \cite{Barke_2015}. The aberration $w_{\mathrm{e}}(r, \theta, 0)$ is composed of normalized Zernike polynomials $Z_{n}^m(\rho, \theta)$ as
\begin{equation}
	w_{\mathrm{e}}(r, \theta, 0) = \sum_{n} \sum_{m} A_{n}^m Z_{n}^m(\rho, \theta)
\end{equation}
where $n$, $m$ are integers and $n>0$ while $m = -n, -n+2, -n + 4, \dots,n-4, n-2,n$. The root-mean-square (RMS) value $w_{\mathrm{e, RMS}}$ of the transmitted aberration is referred to as 
\begin{equation}
	w_{\mathrm{e, RMS}} = \sqrt{\sum_{n} \sum_{m}\left( A_n^m \right)^2}
\end{equation}
subject to the normalization condition. Since Zernike polynomials are defined over a unit circle, the radial coordinate must be normalized as $\rho = r / r_0$ when describing wavefront aberrations across apertures with arbitrary radius $r_0$. 

\begin{figure}
\centering\includegraphics[width=\textwidth]{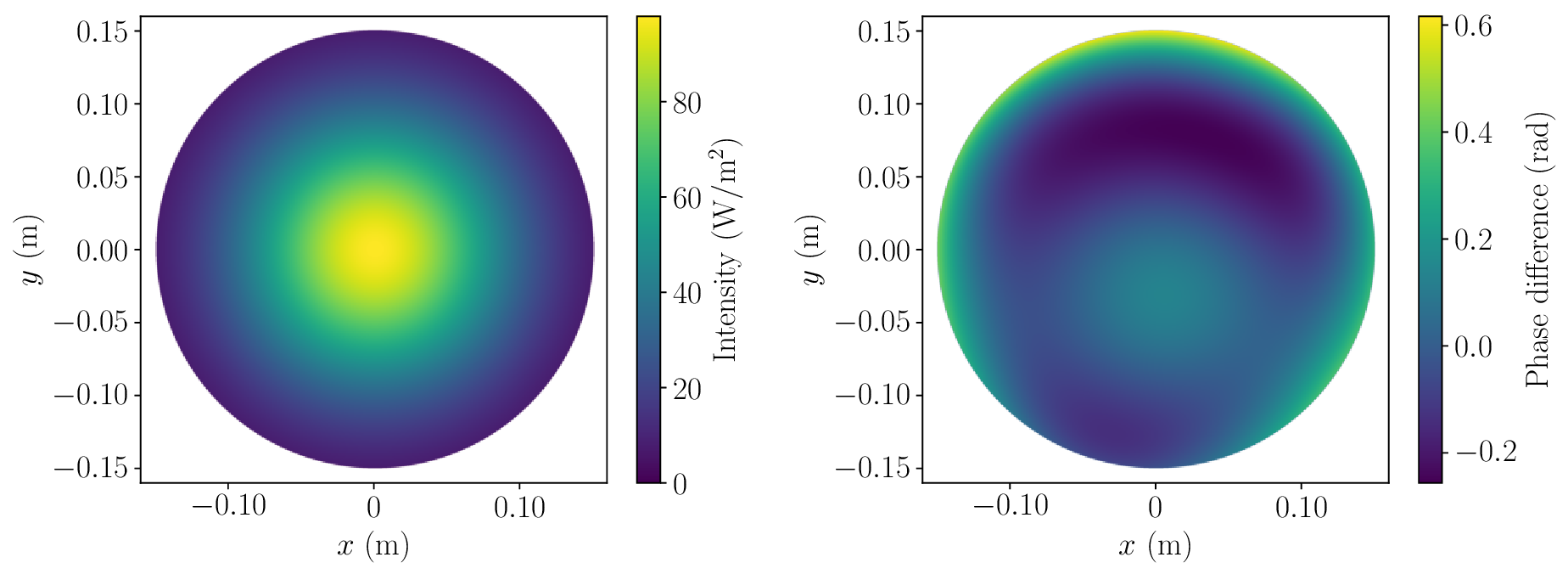}
\caption{The intensity/phase difference distributions of the transmitted beam in the telescope aperture.}
\label{fig:transbeam}
\end{figure}

In this study, the transmitted beam's wavefront aberration is characterized by the first 35 Zernike polynomials, explicitly excluding the piston term because it only corresponds to a global phase shift that does not affect relative wavefront error. Our focus is on aberrations from the 3rd order (defocus) to the 10th order (spherical aberration), quantifying the RMS value of them as an argument of simulations. The total RMS value of higher-order aberrations is constrained to 1 nm to represent roughness, while the first 2 terms, representing tilt, are adjustable via PAAM. The intensity and wavefront error distributions of the transmitted beam are illustrated in Fig \ref{fig:transbeam}.

\begin{figure}
\centering\includegraphics[width=\textwidth]{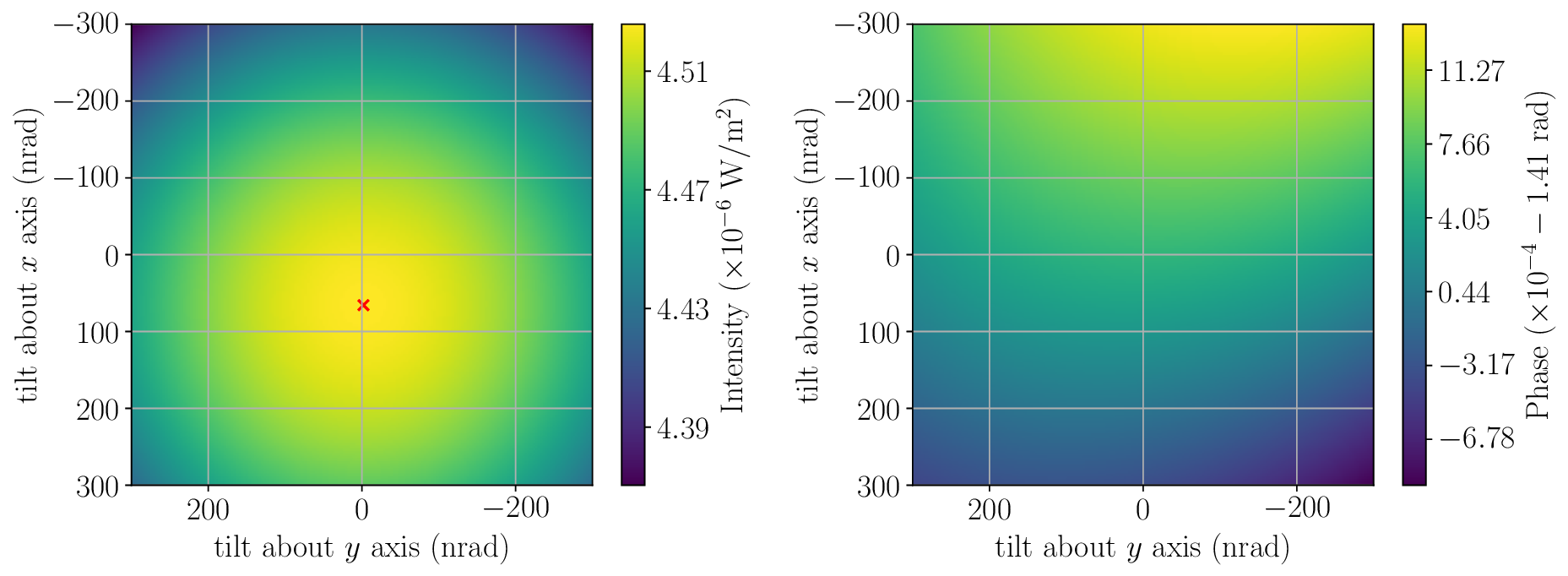}
\caption{The intensity/phase distributions of a transmitted beam in far field and the red cross marks the maximum. Both figures represent positions on the spherical surface using tilt angles about the $x$- and $y$-axes, with sign conventions matching the corresponding Zernike tilt terms.}
\label{fig:farField}
\end{figure}

The numerical solution of Eq. \ref{eq:diffractionEquation} provides full characterization of far-field distributions. Fig. \ref{fig:farField} illustrates the transverse far-field distribution on a spherical surface centered at the TM's CoM, with a radius of $1.732 \times 10^7$ m. The position on the surface in Fig. \ref{fig:farField} is denoted by the angular tilt about the $y$-axis and the $x$-axis relative to the $z$-axis of the beam frame, respectively. Under the influence of transmitted aberration, the maximum-intensity point, marked by a red dot, shifts away from the nominal axis $z$, and the far-field wavefront also exhibits distortion compared to an ideal spherical surface.

\begin{figure}
	\centering\includegraphics[width=0.7\textwidth]{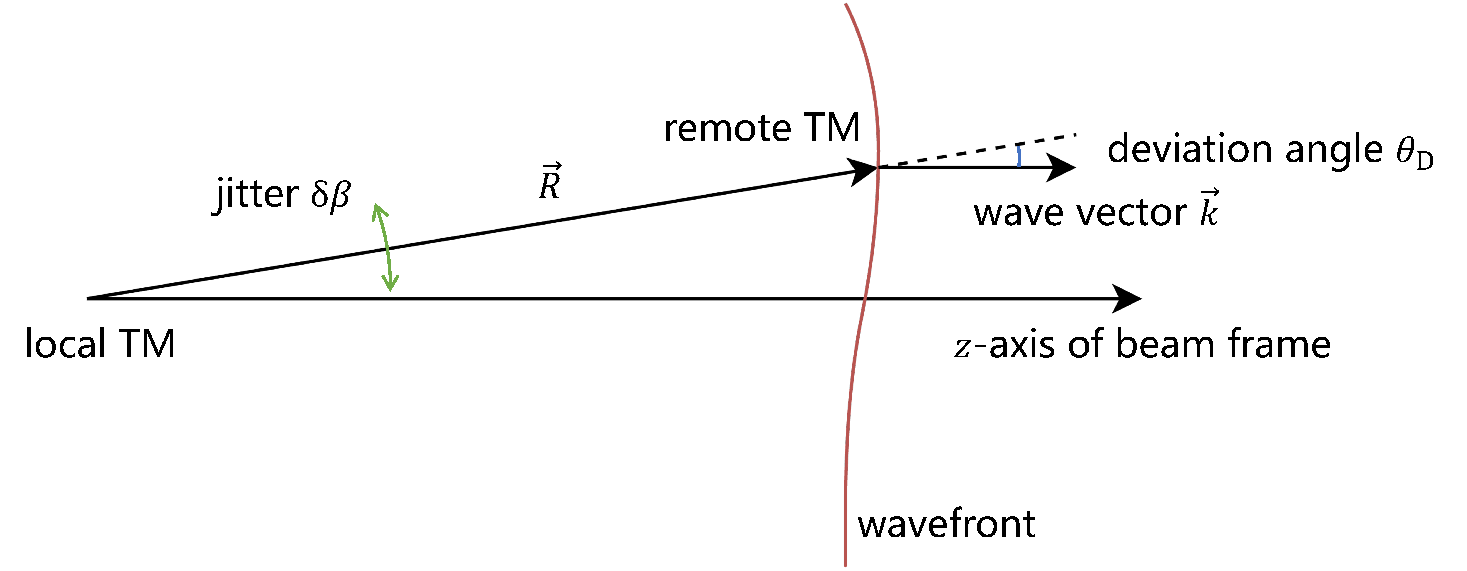}
	\caption{The diagram illustrates the TTL coupling induced by wavefront distortion under beam jitter.}
	\label{fig:ttlPrinciple}
\end{figure}

The TTL coupling coefficient induced by wavefront distortion is a fundamental performance metric for evaluating the wavefront. The TTL coupling effect is concerned with the jitter direction, we adopt the maximum TTL coupling coefficient to evaluate potential measurement errors induced by the wavefront aberration. As illustrated in Fig. \ref{fig:ttlPrinciple}, the gradient of the far-field phase is the wave vector $\vec{k}$, and the presence of transmitted aberration causes the non-zero deviation angle $\theta_{\mathrm{D}}$ between wave vector and the position vector $\vec{R}$. When the transmitter undergoes an angular jitter of $\delta\beta$, the remote TM experiences a displacement $\delta \vec{r}=R\delta\beta\vec{e}_{\beta}$ relative to the beam, where $\vec{e}_{\beta}$ denotes the unit vector along the jitter direction. The resulting phase variation $\delta \Phi$ can then be expressed as
\begin{equation}
	\delta \Phi = \vec{k} \cdot \delta \vec{r} = \left( \vec{k} \cdot \vec{e}_{\beta} \right) R \delta \beta
\end{equation}
 The maximum value of the inner product $\vec{k} \cdot \vec{e}_{\beta}$ equals the magnitude of the wave vector's lateral component, given by $k \cdot \sin(\theta_{\mathrm{D}})$. Accordingly, the maximum TTL coupling coefficient is given by
\begin{equation}
	\left(\frac{\partial L}{\partial \beta}\right)_{\mathrm{max}} = \frac{1}{k} \left(\frac{\partial \Phi}{\partial \beta}\right)_{\mathrm{max}}= R\cdot \sin \left(\theta_{\mathrm{D}}\right) \approx R \theta_{\mathrm{D}}
	\label{eq:deviateAndTTL}
\end{equation}
It follows that the TTL coupling coefficient in a specific direction is a component of the maximum value. Under the PAA compensation strategy of TianQin, the TTL coupling induced by distorted wavefront contributes less total TTL noise but more rate of variation. In light of this, we propose that the TTL coupling coefficient induced by wavefront distortion should not exceed 1 pm/nrad, and its ROC should not exceed 0.08 pm/nrad/day.

\section{Methods for calibration of PAAM}
\label{sec:alignMethod}
To align the transmitted beam with the remote satellite, we define the direction that maximizes the received intensity as a reference. This requires the determination of maximum-intensity direction. For the far-field distribution computed from Eq. \ref{eq:diffractionEquation}, the maximum-intensity direction can be efficiently located via gradient ascent optimization. However, in practice, it is challenging to obtain an accurate measurement of far-field intensity due to various disturbances, including sensor readout noise, transmitter jitter, and the relative movement of the receiver. The acquisition sensor onboard is a primary sensor for measuring the received light power, which is designed to allocate 1\% total optical power collected by the telescope. After the acquisition, the remote satellite is in the beam divergence region, yet may still maintain a certain offset from the maximum intensity. By adjusting the direction of the transmitted beam on the local satellite and monitoring the variations of received optical power on the remote satellite, the reference direction can be further determined and aligned with the remote satellite. This process also constitutes the calibration of PAAM.

\begin{figure}[htbp]
	\centering\includegraphics[width=0.9\textwidth]{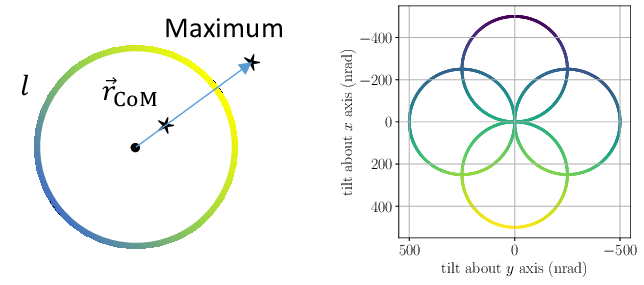}
	\caption{The scanning patterns for searching the peak of intensity.}
	\label{fig:scanpattern}
\end{figure}

As shown in Fig. \ref{fig:farField}, the far-field intensity pattern exhibits approximate rotational symmetry about its maximum-intensity position, which can be exploited to locate this peak. As illustrated in the left panel of Fig. \ref{fig:scanpattern}, a circular scanning trace $l$ lying in an intensity distribution, with the maximum is clearly marked. It can be demonstrated that the "center of mass" $\vec{r}_{\mathrm{CoM}}$ of the field distributed on $l$, the center of $l$, and the maximum point are collinear (see appendix \ref{app:DIDPS}). Specifically, once the CoM is calculated, the direction from the trace center to the field center can be determined. A quatrefoil scanning trace composed of four circular contours, as shown in the right panel of Fig. \ref{fig:scanpattern}, is then proposed. By determining the CoM direction for each contour, the intersection of these directions pinpoints the intensity maximum. We refer to this approach as the direction intersecting (DI) mode.

Alternatively, the symmetry characteristics of the quatrefoil scanning trace can be further exploited to estimate the gradient of the field at the center of the trace (also see appendix \ref{app:DIDPS}). Based on this estimated gradient, a gradient ascent algorithm is then employed to iteratively converge to the intensity maximum. The principle underlying this method is analogous to the differential power sensing (DPS) method \cite{Muller2017}, which is employed to align the beam with the center of the PD on the optical bench. Therefore, this approach is referred to as the DPS mode.

However, the read noise and other disturbances can introduce errors in gradient estimation, which potentially lead to slow convergence or even divergence of the calibration process. To mitigate the impact of such estimation error, we incorporate the Adam algorithm from the field of machine learning, which is specifically designed for finding the maxima or minima of objective functions \cite{kingma2017}. Adam operates by maintaining the exponential moving average of the gradient (i.e., the estimate of the first moment $\vec{m}$) and the component-wise squared gradient (i.e., the estimate of the second moment $\vec{v}$). These moment estimations are used to adaptively adjust the step size during each iteration, thereby enhancing the convergence speed, stability, and robustness (see appendix \ref{app:Adam}).

\begin{figure}
	\centering\includegraphics[width=\textwidth]{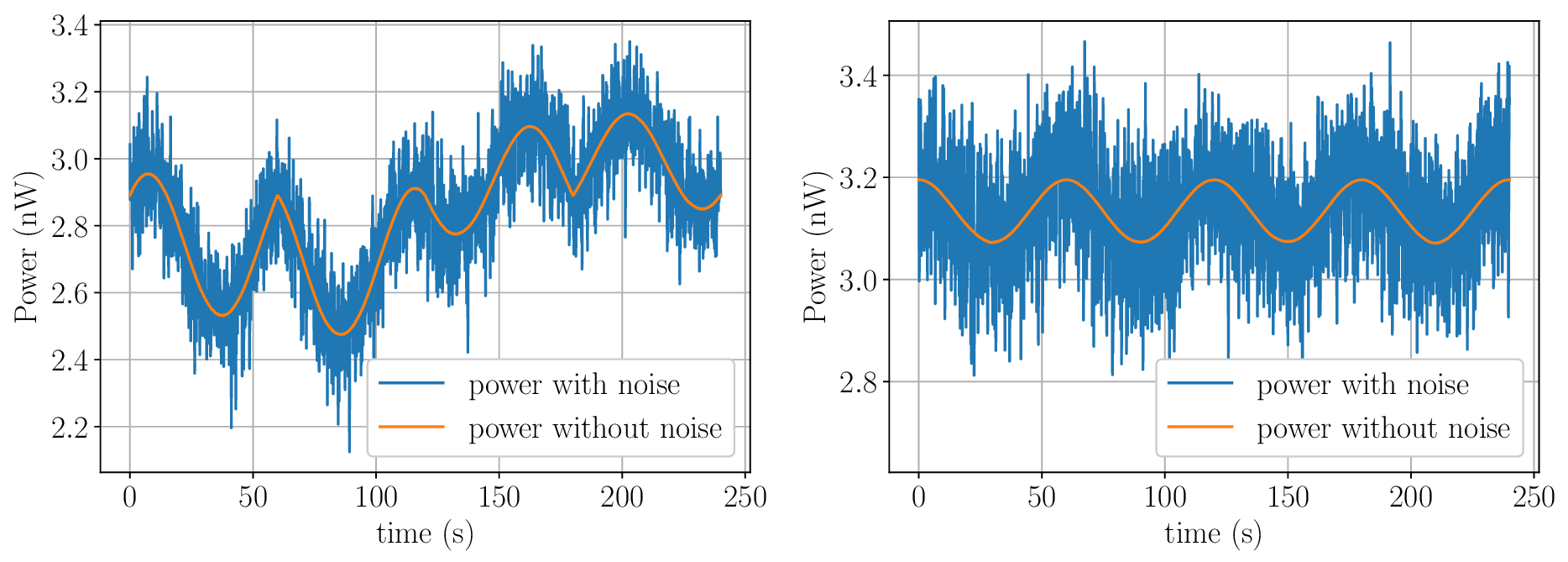}
	\caption{Figures of measured power during a scanning period. The left one illustrates a case with the remote satellite located near the beam edge. The right one illustrates a case with the remote satellite located at the beam center.}
	\label{fig:scanPower}
\end{figure}

We simulated TianQin's PAAM calibration process using these approaches. The simulation primarily involves the following types of noise and disturbances:
\begin{itemize}
	\item The received power readout noise, which is normally distributed with a standard deviation ($\sigma$) of 0.1 nW.
	\item The aberration of the transmitted beam, the RMS value of which is $\lambda$/40.
	\item The pointing bias of MOSA, which is uniformly distributed over the interval [0, 10] nrad.
	\item The jitter of MOSA, which is a noise with the ASD of 5 nrad$/\sqrt{\mathrm{Hz}} \times \left( \sqrt{1 + \left( \frac{f}{60 \ \mathrm{mHz}} \right)^4} \right)^{-1}$ in both $x$ and $y$ directions.
	\item The pointing error of the transmitted beam, which is uniformly distributed over the interval [-0.25, 0.25] nrad in both $x$ and $y$ directions.
	\item The pointing jitter of the transmitted beam, which is a white noise with the ASD of $0.25 \ \mathrm{nrad/\sqrt{Hz}}$.
	\item The position and replacement of the remote satellite, which is derived from the trace shown in Fig. \ref{fig:paaAndTraces}.
\end{itemize}
Under these conditions, the scanning amplitude (i.e., the diameter of the circle in the quatrefoil scanning trace) is selected to 500 nrad, which ensures that measurable power variations can be captured by acquisition sensor whether the remote satellite is located at the beam center or near the beam edge (see Fig. \ref{fig:scanPower}). Each complete scanning period is 4 minutes, with the sampling frequency of 10 Hz. The specific parameters for the Adam method can be found in the appendix \ref{app:Adam}. The left panel of Fig. \ref{fig:moditeration} illustrates the convergence behavior of different approaches under identical conditions. It can be observed that the DI mode exhibits the fastest convergence rate but with relatively poor accuracy, whereas the DPS mode converges more slowly yet attains better accuracy. The Adam method shows the slowest convergence behavior but yields the highest precision.

\begin{figure}
	\centering\includegraphics[width=\textwidth]{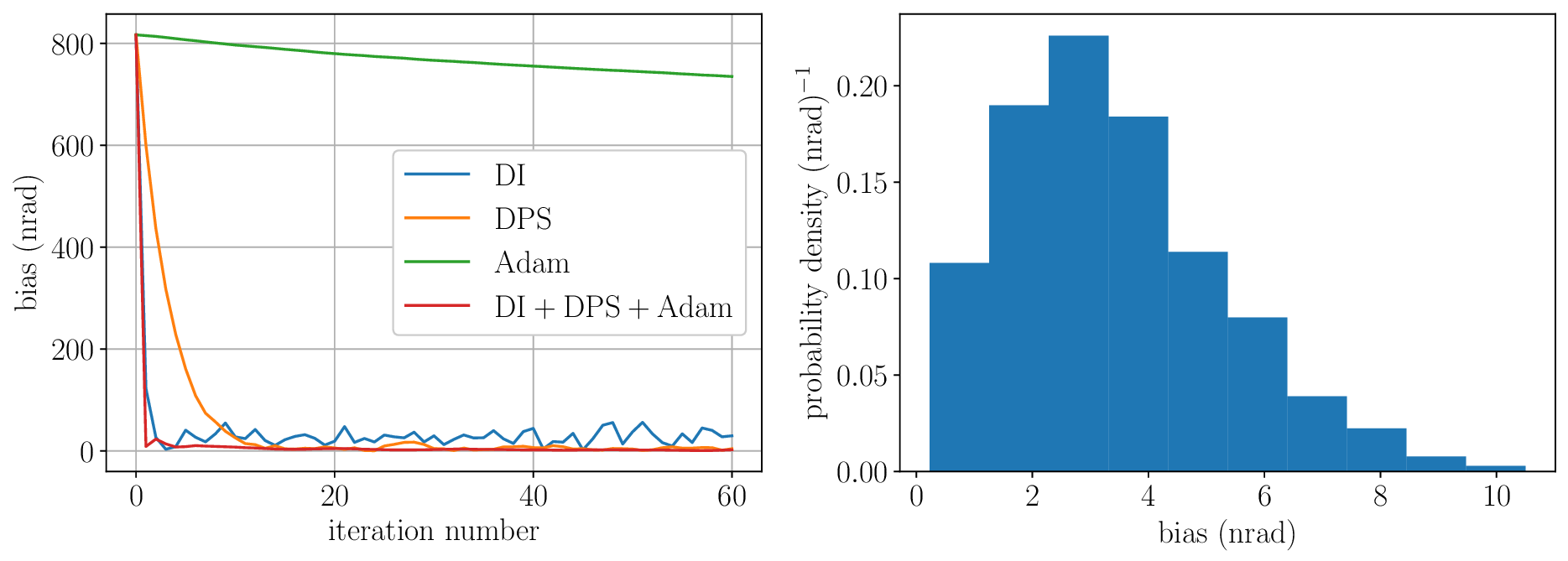}
	\caption{(Left) The figure of convergence properties of each approach. (Right) The histogram of the final bias between the intensity peak and the remote satellite of 1000 Monte Carlo simulations of the alignment process.}
	\label{fig:moditeration}
    \end{figure}

To balance the accuracy and convergence rate, we propose a hybrid PAAM calibration process for TianQin. After the acquisition, the transmitted beam typically exhibits an initial pointing bias, with the receiver positioned approximately 0.7–1 $\mu$rad away from the beam’s maximum intensity. This bias is rapidly reduced to within 100 nrad via 2 scans in DI mode and 6 scans in DPS mode for pre-tuning. Subsequently, the PAAM iteratively performs scans and adjusts the transmitted direction using Adam until the termination criterion is satisfied. For the preliminary configuration, this termination condition is set to a total of 30 scanning iterations, corresponding to an overall duration of approximately two hours. Results from 1000 Monte Carlo simulations (see right panel of Fig. \ref{fig:moditeration}) demonstrate that this process can reduce the deviation between the remote satellite and the maximum-intensity position to below 15 nrad. It can be anticipated that by further optimizing the termination criteria and parameters of Adam, the residual deviation can be further reduced while the time expenditure can be simultaneously decreased.

\section{Alignment strategies for TianQin's inter-satellite beams and feasibility analysis}
\label{sec:simulations}

\begin{figure}[htbp]
	\centering\includegraphics[width=\textwidth]{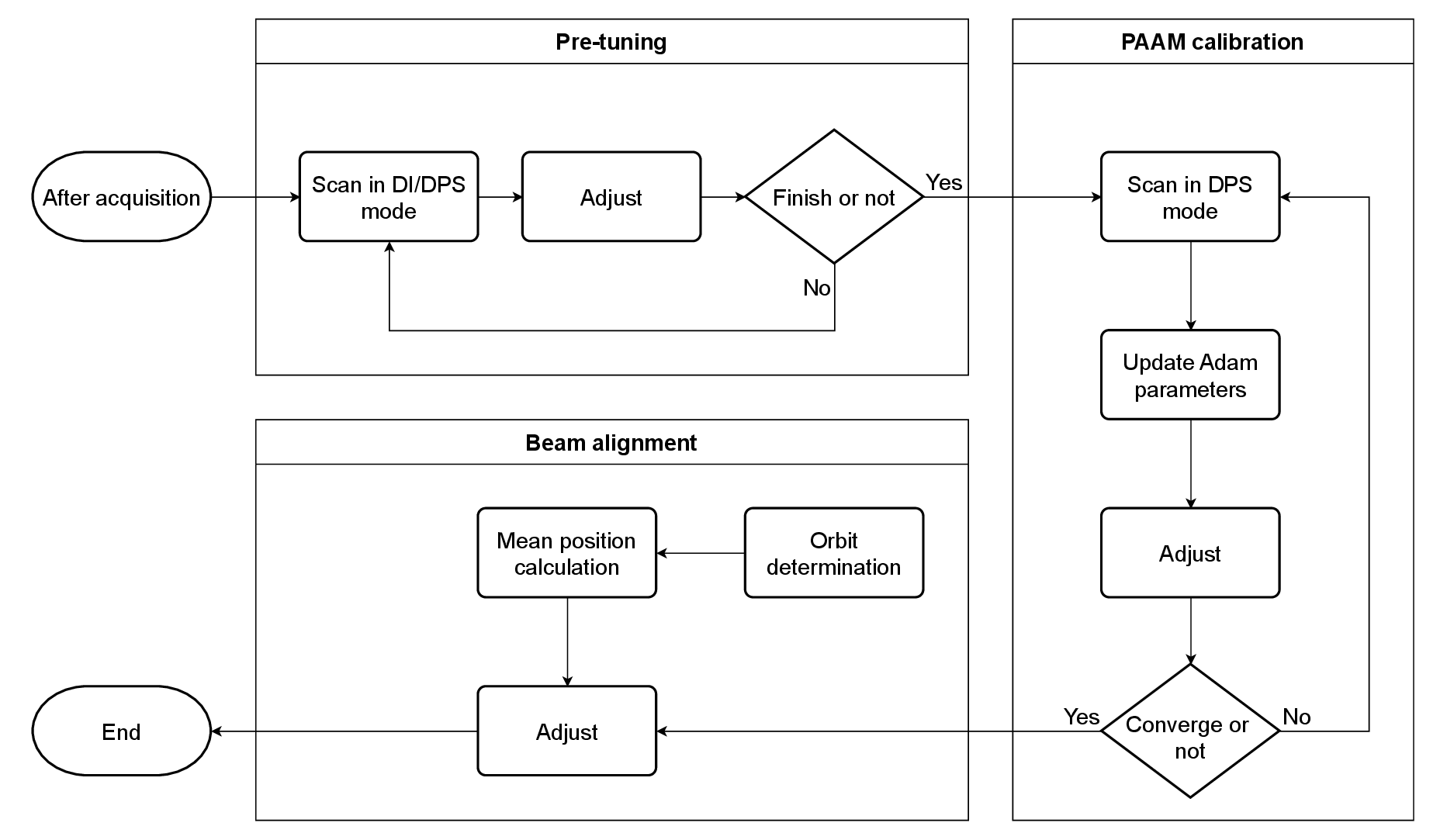}
	\caption{The process of inter-satellite transmitted beam alignment proposed for TianQin.}
	\label{fig:alignprocess}
\end{figure}

We present a set of preliminary alignment strategies for TianQin inter-satellite laser links. For the receiving optical path, each MOSA should be aligned with the incoming light direction, that is, the angle between the wavefront and the QPD plane, which is measured by DWS, is nominally zero. Meanwhile, for the outgoing optical path, consistent with the PAA compensation strategy, the transmitted beam direction is expected to remain stable during science mode. Therefore, the maximum-intensity position of the transmitted beam should align with the mean position of the remote satellite. This requires that, after completing the calibration process for PAAM, the maximum-intensity position must be further adjusted from the remote satellite's instantaneous position to its mean position, which is derived through orbit determination and prediction. The complete process of inter-satellite transmitted beam alignment is shown in Fig. \ref{fig:alignprocess}. In this section, we conduct a series of simulations to examine the feasibility of these strategies.

\subsection{Impact of constellation stability}

Since the mean position of the remote satellite is derived from the orbit determination information, the fidelity of orbit determination can affect the final alignment precision. Based on the orbital simulation, we can discuss the impact of orbital determination errors on TianQin's alignment. The simulation results reveal that the orbital determination errors (the position error in radial direction is $\sim$10 m and the velocity error in along-track direction is $\sim$1 mm/s for per satellite of TianQin) induce a bias in the prediction of the remote satellite's mean position of less than 5 nrad. Synthesizing the calibration capability of the PAAM (Fig. \ref{fig:moditeration}) and the margin, we can draw a conclusion that the total misalignment of the transmitted beam can be kept below 20 nrad.

In general, the remote satellite exhibits motion around its mean position and the stability of orbital directly influences the deviation of the remote satellite between its instantaneous position and its mean position. To assess the impact of orbital stability on the feasibility of the alignment strategy, we performed Monte Carlo simulations to analyze the effects of bias between the maximum-intensity point and the remote satellite. The simulations examined bias magnitudes ranging from 0 to 1000 nrad. For each magnitude, 1000 trials were performed to examine the effects of such bias on the received optical power and the TTL coefficients, incorporating random directions of bias (ranging from 0 to 2$\pi$), random times (corresponding to random orbital positions) and random transmitted wavefront aberrations (RMS = $\lambda$/40). 

\begin{figure}[hbtp]
	\centering
	\includegraphics[width = \textwidth]{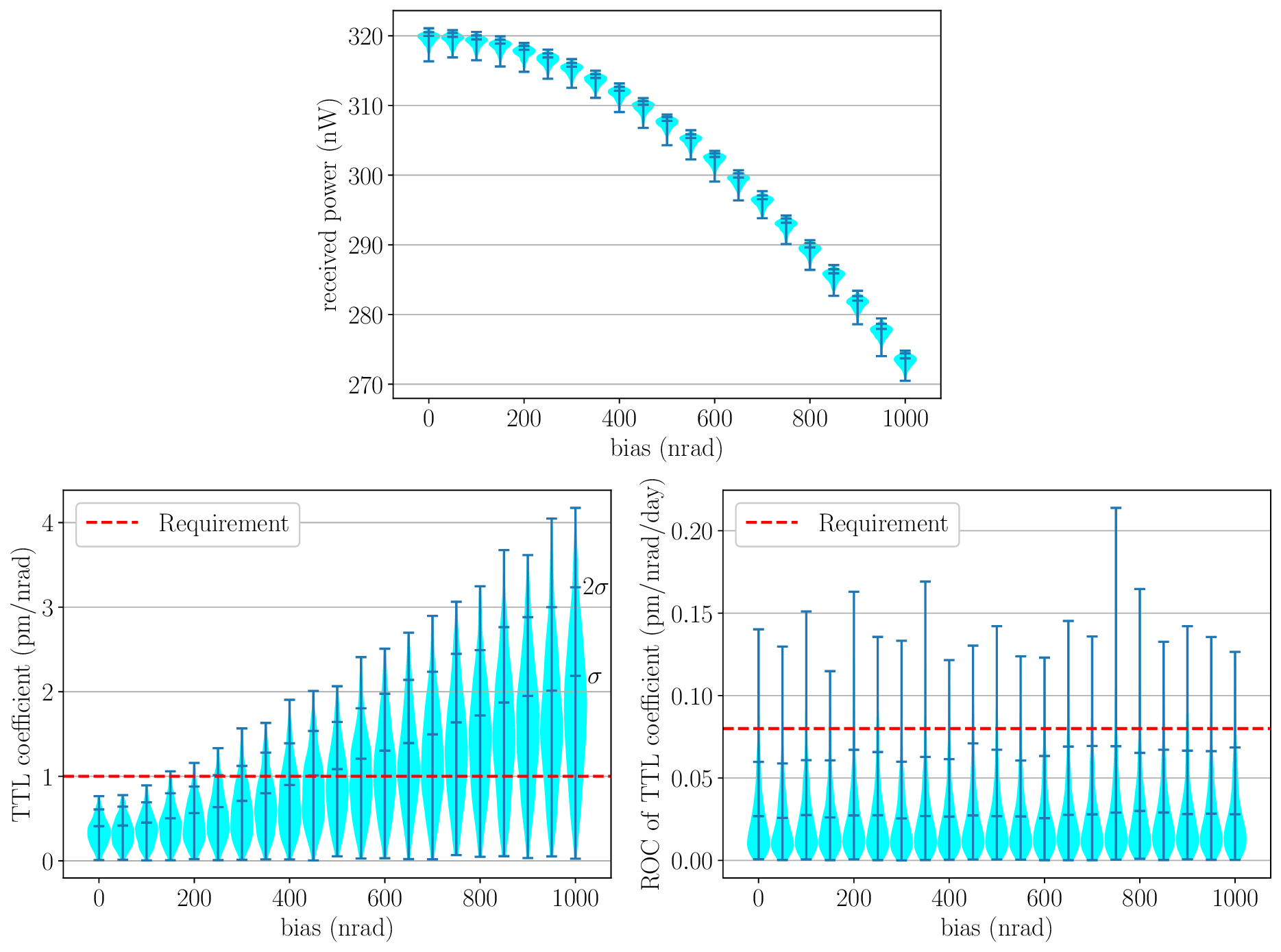}
	\caption{The far-field Monte Carlo simulation results. (Upper) The distribution of received optical power versus misalignment. (Lower left) The distribution of maximum TTL coefficients versus misalignment. (Lower right) The distribution of ROCs of the TTL coefficients versus misalignment.}
	\label{fig:all_vs_bias}
\end{figure}

In Fig. \ref{fig:all_vs_bias}, we employ violin plots to depict the resulting distributions. For each violin plot, the four horizontal bars from bottom to top represent the minimum value, the 68.27\% quantile, the 95.45\% quantile, and the maximum value, respectively. The upper figure demonstrates that the received power declines as bias increases. Furthermore, as illustrated in the lower left panel of Fig. \ref{fig:all_vs_bias}, the maximum TTL coefficient exhibits a discernible dependence on misalignment, with a probability exceeding 95.45\% of remaining below 1 pm/nrad at a misalignment of 250 nrad. The lower right panel of Fig. \ref{fig:all_vs_bias} indicates that ROCs of TTL coefficients follow distributions that are independent of the misalignment magnitude.

The statistical analysis demonstrates that the bias between the maximum-intensity point and the remote satellite has an acceptable impact on gravitational wave detection when it is less than 250 nrad. As the maximum-intensity point aligns with the mean position within 20 nrad, we propose a requirement on TianQin's orbital stability to maintain deviations below 200 nrad between the remote satellite and its mean position throughout an observation period. It should be noted that the requirement is equivalent to the tolerance for the difference between the real-time PAA and the average of PAA. Therefore, this alignment strategy imposes a less stringent requirement on orbital stability compared to prior specifications.

\subsection{Impact of transmitted wavefront aberration}

 As elaborated in Section \ref{subsec:ffsim}, the far-field wavefront exhibits deviation from an ideal sphere because of the existence of transmitted aberration, which may induce the TTL coupling noise. Due to the alignment criterion is the power of received light rather than the wavefront error, after the completion of alignment the TTL coupling noise is remained and may compromise TianQin's performance if the TTL coupling noise exceeds the acceptable threshold of signal post-processing. In this section, we analyze the impact of transmitted wavefront aberrations and try to find out some constrains to accommodate such alignment strategies.
 
\begin{figure}[htbp]
	\centering
	\includegraphics[width=\textwidth]{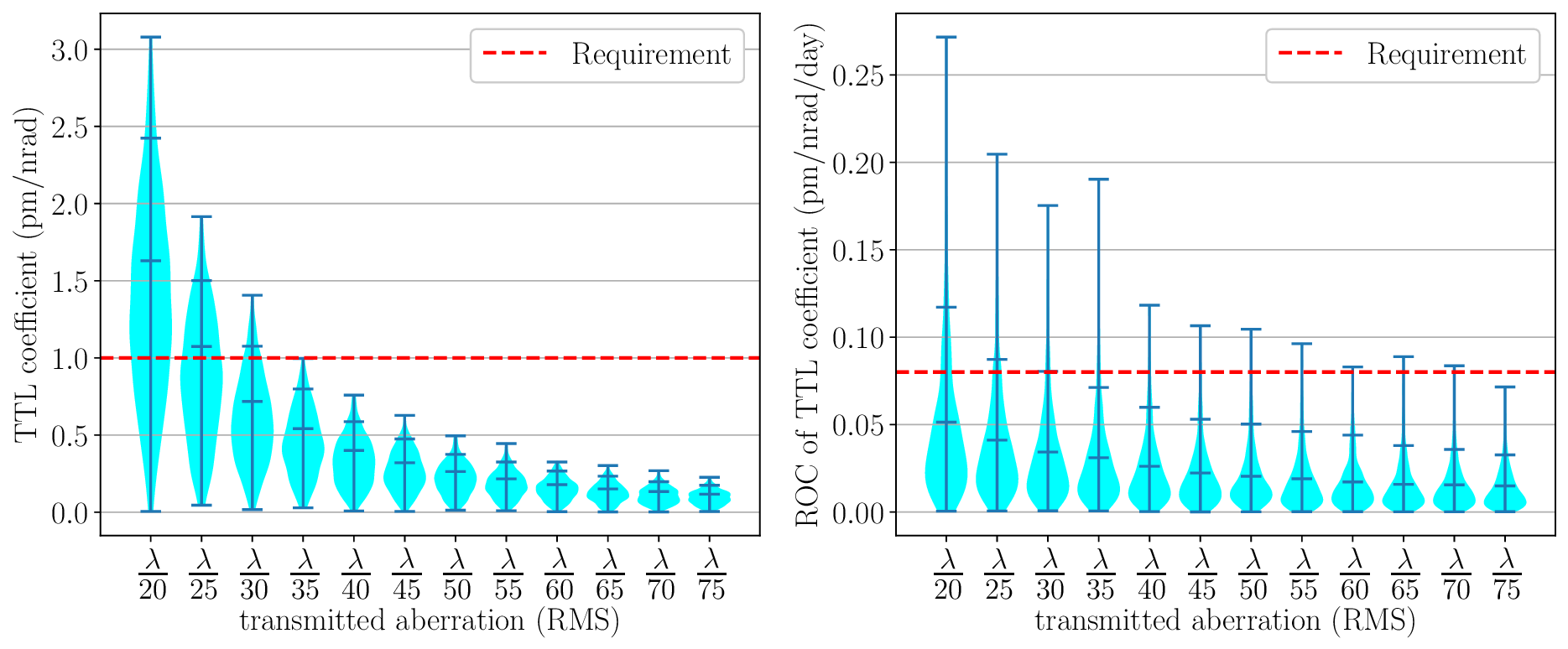}
	\caption{(Left) The distribution of maximum TTL coefficients versus transmitted aberration. (Right)The distribution of ROCs of the TTL coefficients versus transmitted aberration.}
	\label{fig:TTL_vs_RMS}
\end{figure}

 We implement Monte Carlo simulations under different specific aberration RMS values to calculate TTL coupling coefficients and ROCs of them at random times within one observation period, incorporating random pointing direction of MOSA within 10 nrad, random misalignment of transmitted beam within 20 nrad, and random transmitted aberrations. For each RMS value, 1000 simulations are performed. The results, displayed in the Fig. \ref{fig:TTL_vs_RMS}, demonstrate that the distributions of both TTL coupling coefficients and corresponding ROCs diminish with decreasing transmitted wavefront aberrations. To meet TianQin's requirement for inter-satellite TTL coupling noise while accounting for safety margins, optical component fabrication precision, and alignment capability of the optical system, we specify that the RMS of transmitted wavefront aberrations must be below $\lambda$/40. Additionally, the deviation angle derived from Eq. \ref{eq:deviateAndTTL} between the distorted wavefront and the ideal spherical wavefront is $\sim$1 prad, which is significantly below the angular measurement precision of DWS. Consequently, the transmitted wavefront aberrations impose a negligible impact on MOSA alignment.

\section{Conclusion}
\label{sec:conclusion}

This study focuses on alignment strategies for TianQin's inter-satellite laser links, especially on the alignment of transmitted beams. We propose adopting the received optical power as the main criterion of alignment and set the maximum-intensity point of the beam as the reference. Comparing with the stationary point of wavefront which diminishes the TTL coupling noise induced by wavefront error, the maximum-intensity point can be located with less complexity of signal post-processing, less consumption of processing time and higher reliability. The proposed alignment strategy for TianQin is that the MOSA should align with the incident beam while the maximum-intensity direction of the transmitted beam should align with the remote satellite's mean position. To deal with the TTL coupling noise induced by the offset between the wavefront stationary point and the remote satellite, we specify that the deviation between the remote satellite and its mean position must remain below 200 nrad while the RMS value of transmitted wavefront aberrations should remain below $\lambda$/40.

We also propose an iterative adjustment method based on the quatrefoil scanning trace to locate the maximum-intensity direction (i.e., calibrating PAAM), which can complete the fine tuning of the transmitted beam direction in about 2 hours, presumably much quicker than the process of TTL estimation. However, the scanning is implemented via a two-dimensional PAAM. The future work will focus on developing the 2D-PAAM and more detailed signal processing. In addition, experimental validations are necessary to assess the method's reliability and performance under realistic disturbance and noise conditions.

\section*{Funding}

National Key R\&D Program of China (Grant Nos. 2022YFC2204600 and 2020YFC2201202); NSFC (Grant No. 12373116); Fundamental Research Funds for the Central Universities, Sun Yat-sen University.

\section*{Acknowledgements}

The authors thank Yuzhou Fang, Fan Zhu, Lei Fan, Zhizhao Wang, Jinmeng Wang, Gerhard Heinzel, and Jun Luo for helpful discussions and comments.

\section*{Disclosures}

The authors declare no conflicts of interest.

\section*{Data availability}

Data underlying the results presented in this paper are not publicly available at this time but may be obtained from the authors upon reasonable request.

\appendix
\section{Principles of DI and DPS mode}
\label{app:DIDPS}

\begin{figure}[htbp]
	\centering\includegraphics[width=0.9\textwidth]{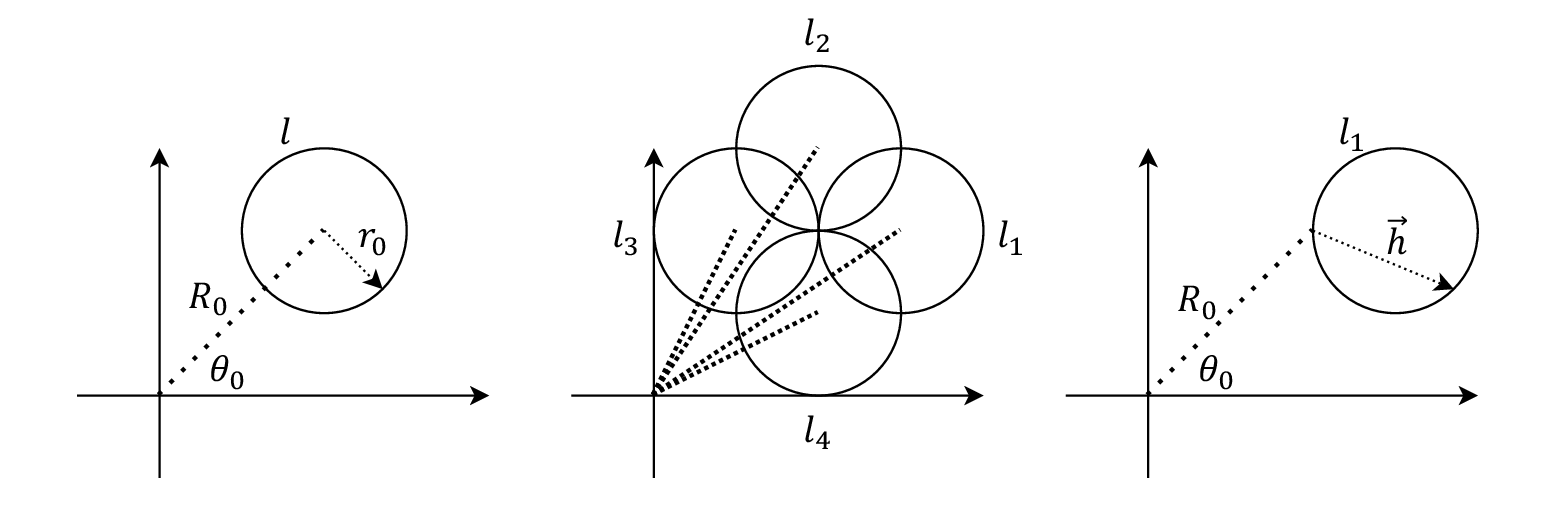}
	\caption{The scanning traces.}
	\label{fig:DI}
\end{figure}

As shown in the left panel of Fig. \ref{fig:DI}, a circular scanning trace $l$ with radius $r_0$ lies within a non-negative definite central scalar field $I(r)$, which is distributed across a plane and centered at the origin without loss of generality. We can perform a calculation on $l$ analogous to determining the center of mass and derive the CoM via
\begin{equation}
	\vec{r}_{\mathrm{CoM}} = \frac{\oint_{l}\vec{r}I(r)\mathrm{d}s}{\oint_{l}I(r)\mathrm{d}s}
	\label{eq:CoM}
\end{equation}
The coordinates satisfy:
\begin{equation}
    \left\{
    \begin{aligned}
        &x\left(\theta^{\prime}\right) = R_0 \cos\left(\theta_0\right) + r_0 \cos\left(\theta^{\prime}\right) \\
        &y\left(\theta^{\prime}\right) = R_0 \sin\left(\theta_0\right) + r_0 \sin\left(\theta^{\prime}\right)
    \end{aligned}
    \right.
\end{equation}
while $\mathrm{d}s = r_0 \mathrm{d} \theta^{\prime}$, and $r = \sqrt{R_0^2 + 2R_0r_0 \cos(\theta^{\prime} - \theta_0) + r_0^2}$. Combining with the Eq. \ref{eq:CoM}, we can deduce that
\begin{equation}
	\vec{r}_{\mathrm{CoM}} = \frac{\int_{0}^{2\pi}\vec{r}g\left(\cos(\theta^{\prime} - \theta_0)\right)\mathrm{d}\theta^{\prime}}{C}
\end{equation}
where $g\left(\cos(\theta^{\prime} - \theta_0)\right) = I\left(\sqrt{R_0^2 + 2R_0r_0 \cos(\theta^{\prime} - \theta_0) + r_0^2}\right)$ and $C = \oint_{l}I(r)\mathrm{d}s /r_0$. It is observed that the vector $\hat{n} = (-\sin(\theta_0), \cos(\theta_0))$, orthogonal to the position vector of the center of trace, simultaneously exhibits orthogonality to vector $\vec{r}_{\mathrm{CoM}}$, which can be derived by
\begin{equation}
	\begin{aligned}
		\hat{n} \cdot \vec{r}_{\mathrm{CoM}} &= \frac{1}{C} \int_0^{2\pi} (\hat{n} \cdot \vec{r}) g\left(\cos(\theta^{\prime} - \theta_0)\right)\mathrm{d}\theta^{\prime} \\
		&= \frac{r_0}{C} \int_0^{2\pi} \sin(\theta^{\prime} - \theta_0) g\left(\cos(\theta^{\prime} - \theta_0)\right)\mathrm{d}\theta^{\prime} =0
	\end{aligned}
\end{equation}
Consequently, we can conclude that the center of the field, the center of the trace, and the center of mass of the trace are collinear.

As seen in the right panel of Fig. \ref{fig:DI}, the average intensity of circle $l_1$ is
\begin{equation}
	\begin{aligned}
		\bar{I}_1 &= \frac{1}{2 \pi r_0} \oint_{l_1} I(r) \mathrm{d}s \\
		&= \frac{1}{2 \pi r_0} \oint_{l_1} I \left( \vec{R}_0 + \vec{h}\right) \mathrm{d}s \\
		&\approx \frac{1}{2 \pi r_0} \oint_{l_1} \left[ I\left( \vec{R}_0 \right) + \vec{h} \cdot \nabla I \left( \vec{R}_0 \right) \right] \mathrm{d}s \\
		&= I\left( \vec{R}_0 \right) + \frac{1}{2 \pi r_0} \oint_{l_1} \vec{h} \mathrm{d}s \cdot \nabla I \left( \vec{R}_0 \right) \\
		&= I\left( \vec{R}_0 \right) + r_0 \frac{\partial I \left( \vec{R}_0 \right)}{\partial x}
	\end{aligned}
\end{equation}
Similarly, we can derive the average intensity for other circles in the quatrefoil line
\begin{equation}
	\begin{aligned}
		\bar{I}_2 &\approx I\left( \vec{R}_0 \right) + r_0 \frac{\partial I \left( \vec{R}_0 \right)}{\partial y} \\
		\bar{I}_3 &\approx I\left( \vec{R}_0 \right) - r_0 \frac{\partial I \left( \vec{R}_0 \right)}{\partial x} \\
		\bar{I}_4 &\approx I\left( \vec{R}_0 \right) - r_0 \frac{\partial I \left( \vec{R}_0 \right)}{\partial y}
	\end{aligned}
\end{equation}
By measuring the difference in average light intensity between opposite-positioned circles, the gradient of the far field can be evaluated as follows:
\begin{equation}
	\begin{aligned}
		\frac{\partial I \left( \vec{R}_0 \right)}{\partial x} &\approx \frac{1}{2 r_0} \left( \bar{I}_1 - \bar{I}_3 \right) \\
		\frac{\partial I \left( \vec{R}_0 \right)}{\partial y} &\approx \frac{1}{2 r_0} \left( \bar{I}_2 - \bar{I}_4 \right)
	\end{aligned}
\end{equation}

\section{Principle of Adam method}
\label{app:Adam}
Given a differentiable function $f(\theta_1, \theta_2, ...)$, where $\theta_1, \theta_2, ...$ represent the parameters, we can iteratively search for its maximum via Adam. The algorithm initiates from a set of arbitrarily chosen parameters, where all elements of first moment $\vec{m}_0 = (m_{1, 0}, m_{2, 0}, ...)$ and second moment $\vec{v}_0 = (v_{1, 0}, v_{2, 0}, ...)$ are set to zero initially. For the $t$-th iteration, the gradient of the function $\vec{g}_t$ is estimated first and updates the first moment and second moment as
\begin{equation}
	\begin{aligned}
		m_{i, t} &= \beta_1 m_{i,t-1} + (1 - \beta_1) g_{i, t} \\
		v_{i, t} &= \beta_2 v_{i,t-1} + (1 - \beta_2) g_{i, t}^2
	\end{aligned}
\end{equation}
where the subscript $i$ before comma indicates the component corresponding to $\theta_i$. The nonnegative hyperparameters $\beta_1$, $\beta_2$, which are less than 1, control decay rates of the history. Since the initial value of $m_{i, t}$ and $v_{i, t}$ are set to zero, they tend to remain biased toward zero during the early stages of iteration. Therefore, bias correction is required as follows:
\begin{equation}
	\begin{aligned}
		\hat{m}_{i, t} &= \frac{m_{i, t}}{1-\beta_1^t}
		\\
		\hat{v}_{i, t} &= \frac{v_{i, t}}{1-\beta_2^t}
	\end{aligned}
\end{equation}   
Finally, the parameters can be updated based on the corrected $\hat{m}_{i, t}$ and $\hat{v}_{i, t}$ like that
\begin{equation}
	\theta_{i, t} = \theta_{i, t-1} + \eta \frac{\hat{m}_{i, t}}{\sqrt{\hat{v}_{i, t}}+\epsilon}
\end{equation}
where $\eta$ denotes the step size baseline, and $\epsilon$ is a small positive constant introduced to avoid division by zero.

In the calibration process of PAAM for TianQin, the two-element function $I(\theta_x,\theta_y)$ represents the received intensity about transmitted direction $[\theta_x,\theta_y]$. To implement the Adam method, the the gradient of $I$ is estimated by scan in DPS mode. After conducting several tests and balancing the convergence speed with stability, we preliminarily selected the following hyperparameter configurations as shown in Table \ref{tab:hyperpara}.

\begin{table}[htbp]
\caption{The hyperparameter configurations in Adam.}
  \label{tab:hyperpara}
  \centering
\begin{tabular}{cc}
\hline
hyperparameter & value \\
\hline
$\beta_1$ & 0.8 \\
$\beta_2$ & 0.999 \\
$\eta$ & 1 nrad \\
$\epsilon$ & $1\times10^{-18}$ \\
\hline
\end{tabular}
\end{table}


\bibliography{refererence}

\end{document}